\documentclass[twocolumn,superscriptaddress]{revtex4-1}
\usepackage{amsmath,amssymb}
\usepackage{graphicx}
\usepackage{color}
\usepackage{subfig}
\usepackage[resetlabels,labeled]{multibib}
\newcites{New}{References}

\begin{document}

\title{Finite-size effects on bacterial population expansion under controlled flow conditions}

\author{Francesca Tesser}
\affiliation{Department of Applied Physics and J.M Burgers Centre for Fluid Dynamics, Eindhoven University of Technology, 5600 MB Eindhoven, The Netherlands}

\author{Jos C.H. Zeegers}
\affiliation{Department of Applied Physics and J.M Burgers Centre for Fluid Dynamics, Eindhoven University of Technology, 5600 MB Eindhoven, The Netherlands}

\author{Herman J.H. Clercx}
\affiliation{Department of Applied Physics and J.M Burgers Centre for Fluid Dynamics, Eindhoven University of Technology, 5600 MB Eindhoven, The Netherlands}

\author{Luc Brunsveld}
\affiliation{Department of Biomedical Engineering, Laboratory of Chemical Biology and Institute of Complex Molecular Systems, Eindhoven University of Technology, 5600 MB Eindhoven, The Netherlands}

\author{Federico Toschi}
\affiliation{Department of Applied Physics and J.M Burgers Centre for Fluid Dynamics, Eindhoven University of Technology, 5600 MB Eindhoven, The Netherlands}


\begin{abstract}

The expansion of biological species in natural environments is usually described as the combined effect of individual spatial dispersal and growth. In the case of aquatic ecosystems flow transport can also be extremely relevant as an extra, advection induced, dispersal factor. There is a lack of reproducible experimental studies on biological fronts of living organisms in controlled streaming habitats. It is thus not clear if, and to which extent, the current theoretical and experimental knowledge on advective-reactive-diffusive fronts for chemical reactions can also apply to the expansion of biological populations. We designed and assembled a dedicated microfluidic device to control and quantify the expansion of populations of \emph{E.coli} bacteria under both co-flowing and counter-flowing conditions, measuring the front speed at varying intensity of the imposed flow. At variance with respect to the case of autocatalytic reactions, we measure that almost irrespective of the counter-flow velocity, the front speed remains finite at a constant positive value. A simple model incorporating growth, dispersion and drift on finite-size hard beads allows to explain this finding as due to a finite volume effect of the bacteria. This indicates that models based on the Fisher-Kolmogorov-Petrovsky-Piscounov equation (FKPP) that ignore the finite size of organisms may be inaccurate to describe the physics of spatial growth dynamics of bacteria.

\end{abstract}

\maketitle

\section{Introduction}

Many biological populations and communities live in liquid environment under the effect of a flow. This occurs both at large scales, for example for aquatic organisms and larvae in rivers and estuaries or marine organisms and plankton in the oceans \cite{speirs2001population, abraham1998generation}, and at smaller scales, for algae in bioreactors down to bacterial infections in human body \cite{singh2012development, costerton1999bacterial}. The study of the expansion of biological species in these environments is relevant for ecology, for example to understand algae blooms or the spread of invasive species \cite{pringle2011asymmetric}, but also, in conservation biology, for the reintroduction and persistence of populations under difficult environmental conditions, like for instance organisms living in the silt of a river \cite{pachepsky2005persistence}. In all cases the complexity of these systems is challenging because of the interplay between living species, with their motility behaviours and their active strategies to persist under difficult conditions and the role of the flow, which is usually both a vehicle for nutrient and the cause of transport of organisms out of their initial environments \cite{mather2010streaming}. Phenomena such as the spatial spreading of populations in new territories and the invasion of new species can dramatically change due to a streaming flow. 

Despite the relevance of these natural phenomena, spatial models for growth, to our knowledge, have never been tested under controlled flow conditions and it is not obvious whether the simple advection-reaction-diffusion scheme, which is usually proposed \cite{ryabovblasius}, can well describe real situations. Microfluidics is currently recognised to be a powerful tool for quantitative studies on microbiological processes and to be the best candidate to control flow at small scales \cite{wu2016nanofabricated, beebe2002physics}. 

In this article we present the design of a dedicated microfluidic device for the growth of \emph{E.coli} bacteria populations under both co- and counter-flowing laminar conditions. We focus our attention to the front dynamics and its advancing speed with the idea of comparing the biological front behaviour to the results reported in literature for chemical species in the context of the advection-reaction-diffusion equation \cite{leconte2003pattern}. 

\section{Spatial growth with advection}

The natural generalisation of classic spatial models for growth in liquid environments is a description in terms of reaction-diffusion-advection equations for the continuum density of organisms, $c(\pmb{x},t)$ \cite{nelson1998non}. In these ecosystems, the spatial dynamics of a population is given by the combination of growth, individual own dispersion, and the transport by the flow as an extra biased migration factor \cite{highRe}: 
\begin{equation}
\frac{\partial c}{\partial t}+\pmb{\nabla}\cdot (\pmb{u}c)=D \nabla^2c+\mu c(1-c).
\label{advection}
\end{equation}
Diffusion models are commonly used for a wide range of living species, such as animals, plants and insects, since their dispersion in many cases is well described by diffusion with a constant coefficient $D$ \cite{pde_in_ecology}. Here, the reaction term is controlled by the logistic growth dynamics with rate $\mu$ towards the stationary concentration, $c=1$, and the advection term contributes for the transport by a flow field, $\pmb{u}(\pmb{x},t)$. 

\begin{figure*}
\centering
\includegraphics[width=17.8cm]{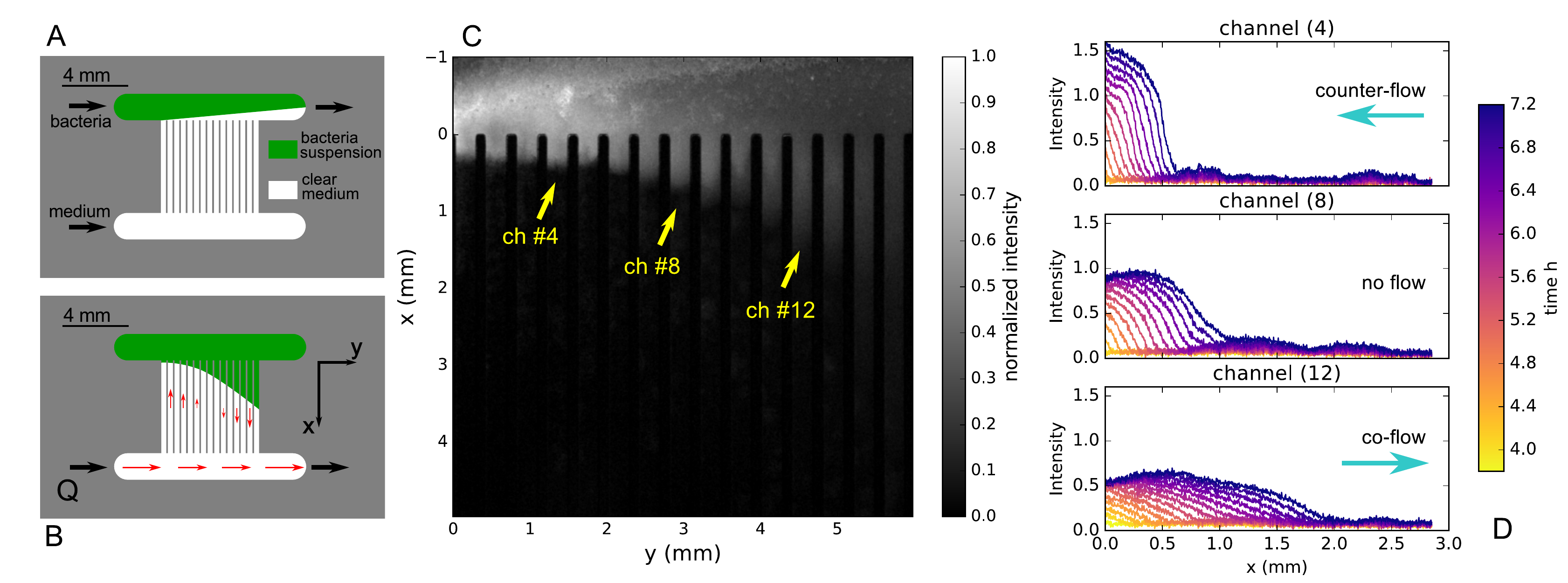}
\caption{(A and B) Sketch of the $15ch$ device and its two modes of operation. (A) Bacteria loading procedure: the bacteria suspension enters from top left and the medium inlet is at bottom left, while both fluids are flowing out from the upper-right outlet. In this way the laminar interface between the two liquids is placed in the upper part of the device as initial state of the experiment. (B) Experimental flow configuration with one inlet for the medium at bottom left and one outlet in the opposite side: this configuration generates circulating streamlines through the channels with varying velocity in opposite orientation. The channel in the middle is characterised by zero velocity for symmetry reasons. (C) Fluorescent intensity map in all channels of the $15ch$ device at time 6.7 hours from the beginning of the experiment. (D) Fluorescent intensity profile along three representative channels, plotted about every 17 minutes, from time 3.8 hours to 7.2 hours, normalised to the maximum intensity in channel 8.}
\label{composite}
\end{figure*}

In absence of any flow, Eq.~\eqref{advection} is the well-known Fisher-Kolmogorov-Petrovsky-Piscounov (FKPP) equation \cite{murray2001mathematical}. For sharp enough initial conditions, like the ones relevant in biology, it admits traveling wave solutions with constant speed $v_F=2\sqrt{D\mu}$ and front width $w$ of order $\sim\sqrt{D/\mu}$ \cite{ebert2000front}. This equation is generally used to describe propagation of population fronts into a homogeneous empty territory. Recently, the invasion speed of ciliate \emph{Tetrahymena} cells has been measured in experiments in a liquid habitat showing a good agreement within the FKPP framework, with the population expansion described in terms of diffusive motility and doubling dynamics \cite{tetrahymena}.
When flows are present, instead, front propagation and transport properties are deeply linked \cite{vulpiani_pre}, and the flow influences not only the expansion process but can also have a dramatic impact on the population chance of survival in the habitat. The solution for Eq.~\eqref{advection} in a one-dimensional homogeneous infinite habitat with constant velocity, of absolute value $u$, is simply given by a change in the reference frame. The two speeds sum up, producing a front propagating with increased speed, $v_f=v_F+u$, if the flow is supporting the growth or reduced speed, $v_f=v_F-u$, in the case of adverse flow. Persistence in the last case is guaranteed for $v_F>u$, otherwise the colony cannot propagate further and is only flushed away \cite{ryabovblasius}. The dynamics gets richer, instead, when the landscape is not infinite  \cite{ryabovblasius} or nonhomogeneous \cite{dahmen2000life,vergni2012invasions} or the flow is nonstationary \cite{speirs2001population} as in realistic situations. 

Several numerical studies have investigated Eq.~\eqref{advection} as models for diluted plankton populations in the ocean under complex turbulent flows, also considering inertial effects \cite{highRe,benzi2009fisher}. Alternative models have been developed to describe situations in which the transport is only partially felt by the populations, like benthic organisms which only occasionally enter into the stream feeling the drift before settling again \cite{pachepsky2005persistence}. 
One experimental application of Eq.~\eqref{advection}, instead, considers the growth of bacteria on a ring-shaped agar landscape rotating with respect to a UV-light pattern to mimic an external advection \cite{lin2004localization}. However, to our knowledge, no systematic and controlled experiments have been performed to quantify the biological front propagation under real flows.

Advection in the case of classical chemical reaction fronts has been instead more studied. The reaction-diffusion equation for this kind of reactions can differ from the FKPP equation in the order of the reaction term. For example, a cubic equation holds for autocatalytic iodate-arseneous acid fronts \cite{edwards2002poiseuille}, but having similar solutions, in particular self-sustained traveling fronts \cite{xin2000front}. Experimental studies based on autocatalytic reactions have been performed in Poiseuille and other laminar flows \cite{leconte2003pattern}, and further experimental investigations have been done with series of vortices \cite{vortices,disordered_vortices} and disordered porous media flow \cite{atis2013,atis2015}. The main benchmark usually consists of the propagation front speed as a function of the mean flow velocity and for autocatalytic reactions under laminar flows there is a good agreement between experiments and advection-reaction-diffusion models \cite{edwards2002poiseuille}. In the supportive-flow regime the reactive front is usually carried by the flow at a speed which is a linear function of the mean flow, while at counter-flow conditions its behaviour appears more diverse. In the Eikonal limit (front width much smaller than the channel width), even if highly distorted by the flow, the front does not slow down but maintains the zero-flow Fisher speed. Instead, in non-Eikonal regimes, it monotonically decelerates reversing the motion for a critical adverse flow \cite{edwards2002poiseuille}. In the case of porous media flows, the front is observed to remain frozen for a wide range of counter velocities before starting to move downstream for relative high adverse flows \cite{gueudre2014strong,atis2015}.

In the case of our experiment we do find a speed up of the front due to a supportive flow and a decrease in the propagation speed with respect to $v_F$ at small counter velocities, but, surprisingly, a stable regime is visible in which bacteria persist growing against the flow at a constant speed, irrespective of the intensity and even for relatively large opposing flows. This result is not explainable in terms of the FKPP equation and has no counter part in the chemical reactions literature.
We will discuss how this result is peculiar of the finite size of bacteria and will conclude that the FKPP is lacking in describing the spatial growth of colonies in this case. This last point shares close similarities with recent non-diffusive models of spatial growth of bacteria colonies on solid substrates \cite{farrell2013mechanically}.

\section{Bacteria front propagation under flow}

In order to perform multiple measurements of bacterial front propagation at different flow conditions in a single experiment we design a fluidic device made of multiple parallel channels with same cross section as sketched in Fig.~\ref{composite}A. 
Most of the sets of experiments reported here have been obtained with a device made of 15 channels ($15ch$), with 300 $\mu$m $\times$ 280 $\mu$m of rectangular cross section and 5.6 mm of length. Furthermore, also the results from a different preliminary device made of 42 channels ($42ch$), with same cross section but 11 mm long, are consistent with the $15ch$ geometry. Results from both the devices are presented here. The devices are made of PDMS and are sealed to glass from one side. A picture of a typical device is visible in the Supporting Information in Fig. S1.
The experiment is characterized by two phases: the loading of bacteria in the device and the actual growth along the channels. For this reason, the device has two inlets and two outlets connected to on-off valves which can be independently set and allow the device to be used according to two different flow configurations. One flow configuration is used for the loading phase and prepares the initial conditions for the experiment by controlling the position of the laminar interface between bacteria suspension and clear medium as sketched in Fig.~\ref{composite}A. The second flow configuration is designed for generating variable flow rates in the parallel channels as depicted in Fig.~\ref{composite}B. The network of flow rates can be solved applying the conservation of mass at each fluidic junction, knowing the fluidic resistance of each branch, and solving for the Poiseuille pressure drop at each node: qualitatively, the pressure drop at the entrance of each channel generates a series of decreasing flow rates in the channels in the first half of the device and increasing flow rates in the opposite direction in the second half of the device. In the case of an odd number of channels, the middle one is characterised by zero flow, for symmetry reasons. This geometry allows to expose the front of bacteria to different velocities both in co- and counter-flow direction within the same experiment. The complete flow field has been solved with a numerical simulation by a Lattice Boltzmann Method (LBM) and verified in the device with Particle Tracking Velocimetry measurements on polystyrene beads diluted in water (results in the Supporting Informations, Fig. S2b). 

With the idea of considering the simplest individual dispersion mechanism, we decide to focus only on thermal diffusion. In the case of non-motile bacteria, an estimate of the diffusion coefficient, using Stokes-Einstein relation for a spherical particle of radius 1 $\mu$m in water at $37^{\circ}$C, gives $D=3\cdot 10^{-7}$mm$^2$s$^{-1}$. 
For this purpose we use \emph{E.coli} $DH5\alpha$, which are known to be poorly motile \cite{wood2006motility} and non-motile on agar \cite{hallatschek2007genetic}. At the beginning of the experiment bacteria are deposited by flow in the upper side of the device, as initial condition, where they diffuse and start to duplicate expanding along the channels. As visible by bright field microscopy, no motility is exhibited at this stage, however a very small fraction of bacteria appears to swim after two and a half hours from the beginning of the experiment and deposit at the walls, also much further than the main front position. This phenomenon, in principle undesired, due to the nonhomogeneous behavior in the population sample, has the effect of limiting the duration of the main front detection in time. However, no swimming motility is visible for the bacteria which compose the main front tracked in our experiment.
Considering a duplication time for bacteria at $37^{\circ}$C equal to $T=38$ min$=\ln{2}/\mu$ \cite{hallatschek2007genetic} and the diffusion as estimated above, the expected Fisher speed is $v_F\approx2\cdot 10^{-2} \mu $m/s and the front width of order $w\sim30 \mu$m.

The bacteria were genetically modified in order to be fluorescent and to be detected by camera equipment and appropriate filters (see Appendix \ref{app3}). The fluorescent intensity map is captured along the channels at intervals of times of approximately 6 minutes. An example of such an intensity map is displayed in Fig. \ref{composite}C. A variable amount of time is needed by the front to reach the entrance of the channels ($x=0$ in Fig. \ref{composite}C and \ref{composite}D), from this moment the front is detected for a period of time of approximately 3 hours (about 8 hours in total from the beginning of the experiment). After this time, the motion of the front gets usually disturbed and then hidden by the background homogeneous growth given by the small fraction of swimming bacteria, as mentioned above.
To obtain quantitative information on bacteria density, the proportionality between the intensity of fluorescent light collected by the camera and the bacteria density of reference samples was validated under the same optical conditions for a significant range of bacteria concentrations (see Fig. S3a in the Supporting Information). The intensity signal is then integrated along the transversal direction in each channel in order to express the front profile in the direction parallel to the flow as shown in Fig. \ref{composite}D for three representative channels.

\begin{figure}
\centering
\includegraphics[width=.9\linewidth]{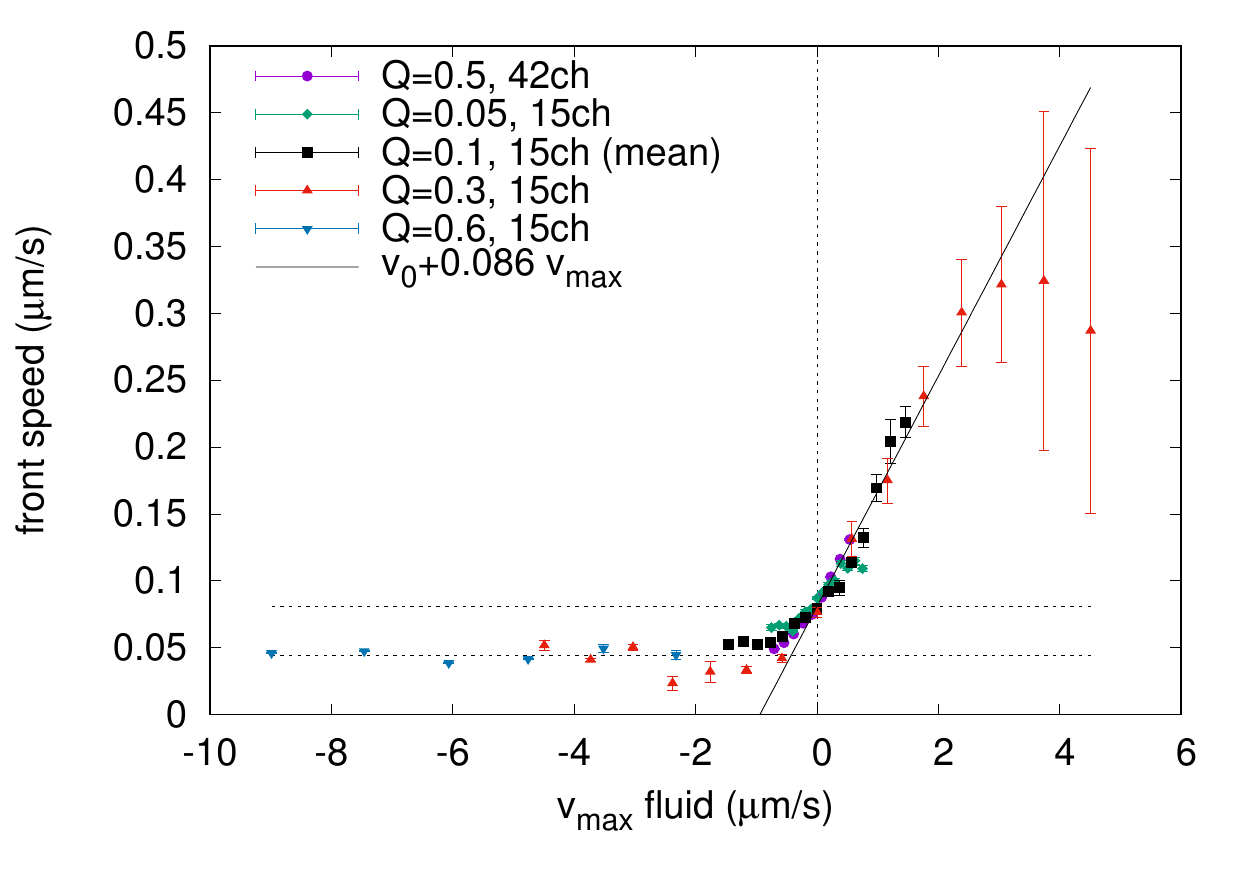}
\caption{Front propagation velocity of bacteria population as a function of the maximum centreline velocity, (black) squares are mean values of four identical experiments, while the other sets of data refer to single experiments. Different medium flow rates Q ($\scriptstyle{\mu}$L/min) and different devices, $15ch$ or $42ch$, span different velocity ranges. While error bars are small in the upstream case, they become in general bigger for the downstream case, due to a less defined front in the second case. Dashed horizontal lines refer to the estimated zero velocity front speed $v_0$ and plateau velocity $v_{min}$, the continuous line is a linear fit on the positive fluid velocity $v_{max}>0$ and corresponds to a slope $v_f=v_0+0.086v_{max}$.}
\label{front_speed}
\end{figure}


\section{Results and Discussion}

The main result of this experiment concerns the velocity of the population fronts along the channels. 
First of all the existence of traveling waves propagating with constant speed along the channels is confirmed, as visible in Fig. \ref{composite}D, for which it is possible to define a front speed (see Supporting Information for details). The results concerning the front speed are summarised in Fig. \ref{front_speed}, as a function of the centreline velocity, $v_{max}$, of each channel. These values of the flow are the ones obtained from the LBM simulations. The experiment is reproduced at different inflow rates $Q$ in order to span a wide range of flow speeds. Moreover, for $Q=0.1~\mu L/\min$ and the $15ch$ device, four identical experiments are repeated in order to analyse the intrinsic variability of the process and to improve the estimation of the output values by averaging over independent tests. In this case only the mean value is shown instead of single measurements. The errors on single measurements are defined based on the technique for speed extraction (see Supporting Information), while the errors on the mean values are obtained by usual error propagation.  
The fast co-flow cases are difficult to analyse in terms of sharp fronts propagating with a defined velocity. Indeed, in this case, the growth process appears to be driven by single individuals which are resuspended and carried downstream at early times and dramatically diluting the front. It is not possible to quantify this growth regime in terms of a collective front propagation on these time and length scales. This extreme co-flow regime is left out of the purpose of this paper and the design of the channel and acquisition procedure are optimized for the intermediate regimes. This choice implicates gradual increase in the error bars in the fast supportive flow regime.
In the set of our experiments, a direct measure of the bacteria propagation speed in liquid medium at zero velocity is available, which results in $v_0=(0.081\pm0.001)~\mu$m/s. It is of the same order of magnitude but a factor four larger than the Fisher speed prediction, $v_F$.
Fig. \ref{front_speed} shows that the front speed varies with the fluid velocity, propagating faster for a flow supporting the growth and reducing for a flow in the opposite direction. In particular, in the co-flow regime the front motion can be explained as an advected front, propagating at a speed which is given by the growth speed at zero velocity, $v_0$, plus a contribution proportional to the fluid velocity. The slope obtained from a linear fit, $s=0.086\pm0.003$, is much smaller than one, which means that the front is not carried downstream by the maximum fluid velocity nor by the mean fluid velocity but by a much smaller velocity. This is consistent with sedimentation of the bacteria used, since they are slightly denser than the medium, combined with a parabolic profile of the flow in the channel, which makes the bacteria experiencing a reduced velocity, closer to the wall. The effective velocity which is relevant for the front speed-up for these data can be estimated from $s$ to be at distance $d\ll h,d\approx hs/4\approx 6 \mu$m from the wall, where $h=280 \mu$m is the depth of the channel and a parabolic profile is assumed along the vertical direction $z$ with no-slip conditions at the walls $z=0$ and $z=h$, and $v_{max}$ as centreline velocity at $z=h/2$.
On the negative axis in Fig. \ref{front_speed}, instead, the front speed is reduced with respect to $v_0$ but, remarkably, does not show a dependency on the flow intensity over a wide range of counter-flow velocities. A plateau velocity appears indeed, where the bacteria front propagates at a constant positive speed independent of the flow velocity. The average of the data in this regime gives $v_{min}=(0.044\pm0.002)~\mu$m/s, which is smaller than $v_0$.  
The experiment is designed in a way that negative population front velocities are not measurable, since they would never enter the region of interest where the acquisition takes place. Anyhow the output of an eventual negative speed experiment would have given no front detection, while a very clear signal, with accurate speed, is always visible at relative high counter-flowing velocities. 
Independent observations in bright field (see Appendix \ref{app4}), of this particular counter-flow regime, show that the population is expanding in its leading part as a monolayer at the level of the bottom wall. From this observation one can inferred that boundary conditions are extremely important to sustain the growth upstream, both because the velocity is zero at the wall and because of eventual interaction, i.e. stickiness, between cells and walls. However it is not clear how this growth speed at the wall differs from the one at zero flow. An explanation is suggested below.

Interestingly, plateau regimes exist also for autocatalytic reactions, but only at zero velocity \cite{atis2015} or at the Fisher velocity for channel flows in the Eikonal limit (front width much smaller than channel width) \cite{edwards2002poiseuille}.  
It can be deduced, therefore, that the presence of two characteristic different velocities, the one at zero flow and the minimum one at strong enough counter-flows, indicates that the growth process is somehow different in the two cases. A possible interpretation of the results considers the fact that at the wall, and with strong opposite flow, any diffusion motion is suppressed and the only propagation possible is the duplication process and the cumulative mechanical forces between individual bacteria. At small velocities, instead, or without any flow, the contribution by the diffusive dynamics, the usual Fisher speed $v_F$, adds up to that minimal growth process $v_{min}$ producing a front which is faster than the growth on a substrate and faster than the Fisher speed alone, $v_0=v_F+v_{min}$.

\begin{figure}
\centering
\includegraphics[width=.9\linewidth]{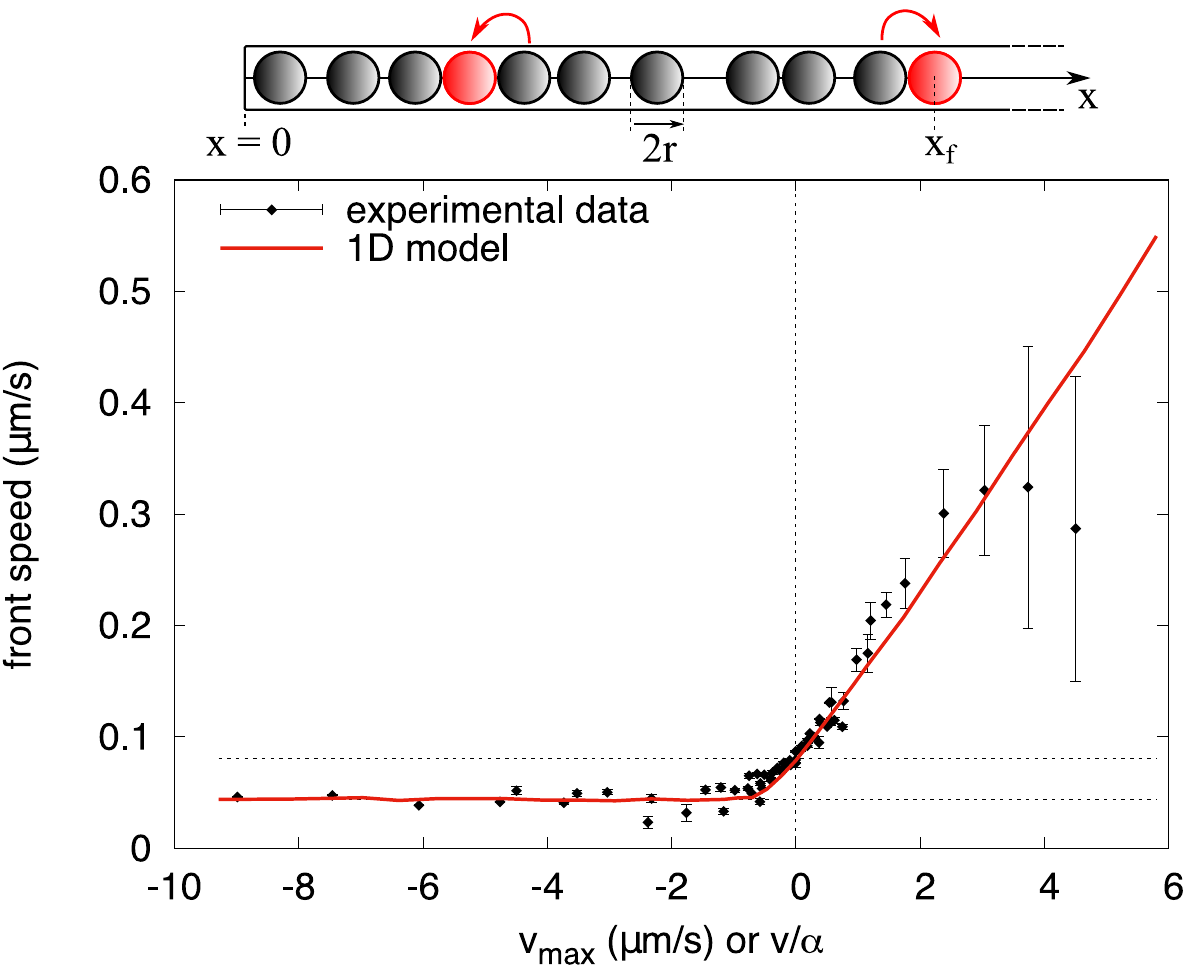}
\caption{(Top) Scheme for the 1-dimensional growth model of hard impenetrable spheres of radius $r$ on a semi-infinite space with a wall at $x=0$. Every particle undergoes a stochastic drift-diffusion process and gives birth to a new particle next to it (in red/light gray) at constant rate only if there is space. The duplication processes are indicated by the arrows.The front is defined as the position of the most advanced particle. (Bottom) Comparison between the experimental data of front speed of Fig. \ref{front_speed} and the numerical front speed from the 1d hard beads model, as a function of drift $v$. Parameters for the model have been fixed in order match the two velocities of the experiment, $c_{min}=v_{min}$ and $c_{0}=v_{0}$. $D=0.0036$, $\mu=0.5$, $r=0.088$ and time step $dt=0.05$. One more parameter is free to match velocity in the channel $v_{max}$ with the drift $v$ in the model $v=\alpha v_{max}$, with $\alpha=1/s$.}
\label{model}
\end{figure}

In order to validate this hypothesis we developed a simple model of hard (impenetrable) spheres in one space dimension which undergo a drift and diffusive dynamics with duplication as sketched in Fig. \ref{model} (see Appendix \ref{app4} for details). Usual models for spatial growth, as the FKPP equation or discrete models as the stepping stone algorithm \cite{korolev2010genetic}, assume a growth contribution which acts only locally and diffusion or migration as spatial phenomena. In this model, instead, also a non local effect is introduced for the birth event, given by the excluded volume of the spheres. The concept of ``motion by growth'' is not new, as it is known that in very narrow channels, where bacteria motility vanishes, bacteria cells can still pass through by growth and division \cite{mannik2009bacterial}. In the model that we propose, it is the diffusive motion which vanishes due to the strong counter-drift, which packed the spheres at the maximum density, and the minimal growth speed is the only mechanism left. In Fig. \ref{model} the front speed from our one dimensional model are compared against our experimental data of Fig. \ref{front_speed}. This is done by tuning the parameters $D,\mu$ and the radius of the spheres, $r$, in the model to match the two velocities, $v_0$ and $v_{min}$ of the experimental data, and rescaling the drift of the model, to match also the slope of the experimental data in the supportive regime. 

So far the discussion focused on the front speeds, however the propagation dynamics is also characterised by the front width and the density at the stationary state.  As visible in Fig. \ref{composite}D, and also confirmed by the data from other channels, the front appears to be sharper and denser in the counter-flow condition, while in the supportive case it looks systematically more dilute and broader (more detailed analysis is found in the Supporting Information in Fig. S5). This observation is in agreement with the picture of the one-dimensional model, by which the density of beads is systematically higher for negative drift velocities, where they become highly compact, while positive drifts tend to dilute the front. However, the comparison is only qualitative. For these observed quantities the transversal dimension and the vertical variation of the flow field are expected to be relevant as well and, probably, a more microscopic information on the bacteria body distribution would be needed in order to improve the model.

\section{Conclusions}
By the use of a microfluidic device we have performed experiments of bacterial colony growth under controlled flow conditions, observing traveling waves and measuring their propagation speed. In addition to a direct measure of the propagation speed at zero flow in liquid medium, we have observed the influence of the flow both in opposite and supportive directions. Surprisingly, a well visible plateau regime can be observed at non-zero front velocity, for the high counter-flow case, which leads us to consider two distinct contributions to the spatial dynamics: the Brownian diffusion on one side and the duplication of finite-size organisms on the other side. In the case of non-motile bacteria in liquid medium, these two mechanisms appear to contribute approximately with the same order of magnitude, so that both effects are visible, but, in general, one of the two contributions can be negligible for other living or chemical species.
Indeed, in the case of chemical species, the local nature of the reaction term is legitimate down to the atomic scales, while in the case of cell size, the front velocity is dominated by the division replication process when the diffusive dynamics is suppressed (e.g. on solid substrates or under strong counter-flow conditions)

For the growth of bacteria in liquid medium, where both Brownian dynamics and advection take place we have suggested a simple model to capture the role of advection and diffusion for finite size duplicating beads. Even though very simple, the model includes the key essential features.  
In principle, the one-dimensional model can be extended by considering other dimensions, or introducing further details like the existing models for growth of bacteria on substrates, which solve the mechanical forces between bacteria allowing also some compressibility on the individual bodies \cite{farrell2013mechanically}.

In conclusion our experiments indicate that advection-reaction-diffusion equations are useful to describe the dynamics of populations growing under flow but, in some circumstances, the FKPP equation needs to be extended to account for finite-volume effects. In particular, these finite-volume effects seem to be relevant when the diffusive dynamics is not dominant, like in the counter-flow regimes where it can be suppressed by advection.
Additional investigations taking into account the complete flow field with its vertical profile, systematic analysis on the effect of varying the shear rate with similar centerline velocities by changing the channel geometry, can also provide additional understanding on the population growth dynamics under flowing conditions.


\appendix

\section{Bacteria strain preparation} 
DH5$\alpha$ bacteria were genetically modified to contain the pHT584 plasmid \cite{ANIE:ANIE200904413}, coding for the expressing of monomeric Yellow Fluorescent Protein (mYFP) out of a pTWIN vector with Ampicillin resistance. The bacterial strain was cultivated overnight in 6mL of growth medium (LB) supplemented with Ampicillin. 2 mL of the bacteria culture was then diluted in 2 mL of fresh medium. Estimation of the bacteria density at this step is $1.5 \cdot 10^{10}$ bacteria/mL, obtained by single colony counting procedure. Bacteria are loaded in the PDMS device using a BD plastic 3 mL syringe, controlled by a KDS Legato 180 syringe pump at flow rates set between 4 and 8 $\mu$l/min for 2 h.

\section{Microfluidic device and experimental protocol} 
The device is made of PDMS and is fabricated following standard soft-lithographic procedures and sealed on a glass slide \cite{soft-lithography}. A new device is fabricated every time for a new experiment, since it is not possible to clean it completely after being used with bacteria. 
The experiment requires an accurate flow control at very small velocity scales, of order of $\mu$m/s and smaller, and also no flow in some parts of the device for a long period of time, of order of 5-10 hours. Because of PDMS permeability to water vapour \cite{permeation-driven}, which is not negligible at these velocity rates and time scales, the device is put under water at 50$^{\circ}$C for minimum 1 hour before placing under the microscope, to saturate the PDMS with water. Then it is immersed inside a water bath for all the duration of the experiment to completely remove residual flows otherwise given by evaporation through the walls. The temperature control is provided by a hot plate in contact with the water bath set at the temperature of 43$^{\circ}$C in order to maintain the bath and the PDMS at 37$^{\circ}$C, optimal temperature for bacteria growth. The temperature calibration has been performed using a probe. The inlets and outlets of the device are connected with 90-degrees bent metal connectors to soft tubing (0.5 mm ID) to 4 independent shut-off PEEK valves, in order to change from loading bacteria configuration to the experimental flow condition. 
The syringe for the medium is a 10 mL Hamilton glass syringe controlled by a separate KDS Legato 180 syringe pump. The medium is degassed with Biotech Degasi Classic before the use, then it is flushed in the device for 30 minutes to replace the water and to provide an homogeneous concentration of nutrient. Then the bacteria syringe is connected and the bacteria loading procedure can take place. The medium and bacteria syringe pumps are synchronised, so that they ramp down automatically after 2h to the final flow rate, minimising the disturbance to the bacteria interface. Finally the valves are manually changed to the final flow configuration.

\section{Image Acquisition}
\label{app3}
Optical access to the device occurs through the glass by use of an upright fluorescence microscope (Olympus BX61) coupled to a CCD ATIK-4000 camera. The appropriate excitation and emission of fluorescent bacteria strain is provided with the filter cube MYFP-HQ by Olympus (excitation BP 490-500 nm, emission BP 515-560 nm)  and light is focused and collected by a 2.5x magnification objective, which gives a field of view of 6x6 mm$^2$ (resolution $\sim$3$\mu$m/pixel). In general acquisition of fluorescent images is done at intervals of 343 seconds with an exposure time of 30 seconds. Bright field mode has been used in the experiment to check the initial state of the experiment and to monitor the leading front progression at high counter-speed with 10x and 20x magnification objectives and a THORLABS camera.

\section{1D growth model of hard beads}
\label{app4}
The model considers a number of one dimensional hard beads of diameter $2r$ living on a one dimensional continuous semi-infinite habitat $x=[0,+\infty[$. The dynamics of a single bead is described by a stochastic displacement $dx$ given by a constant drift $v$ and diffusion $D$:
\begin{equation}
\scriptstyle{dx=v dt+\sqrt{2D}dW(t)}
\end{equation}
with $dW$ the increment of a Wiener process $W(t)$ with unit variance: $\langle dW(t)\rangle=0$ and $\langle(dW(t))^2\rangle=dt$. The beads interact by excluded volume, in the sense that they are impenetrable and cannot overlap. Since the system is confined in 1 dimension, the order is also necessarily preserved, as in the single file dynamics and this has important consequences for the statistical behaviour of the constrained beads in the bulk \cite{sabhapandit2007statistical}. Moreover growth is also implemented: each bead can give birth to an identical bead of the same size next to it, left or right, at a rate $\mu$ and only if there is space. The actual probability for a particle in $x_i$ to give birth on the right in the time interval $dt$ is then $\mu dt/2$ multiplied by the probability of the space between $x_i+r$ and $x_i+3r$ being empty. A wall is placed at $x=0$ and it has the same effect of a fixed bead at $x=0-r$. The dynamics advances at intervals of time $dt$ in which beads displace and have chance to duplicate in a random order, the overlap between beads is treated as a kind of inelastic interaction: the particle of interest is displaced as much as there is place and then, in case of overlap, it is placed next to the neighbouring bead, which means at the minimal distance $2r$ from the its centre. Note that here the advection is not implemented as a pure transport on the whole population, but as a biased drift on each individual movement.
The front is defined as the position of the most advanced particle, $x_{f}(t)$, and the front speed as the derivative, $c=dx_{f}/dt$. It is numerically estimated as the slope of linear regression on values of front position on time, averaged over independent realizations. Depending on the direction of the drift $v$, the dynamics is classified in three classes: the no-flow $v=0$, the counter-flow $v<0$ or the co-flow $v>0$ case. At zero velocity the front speed is given by a constant, plus a term which scales as the Fisher speed: $c_0 (D) = c_{min} + A\sqrt{D}$, however even in the limit $D\rightarrow 0$ the front can proceed at its minimum speed, which is given by duplication of the most advanced particle, at velocity given by one body size $=2r$ per twice the duplication time $\tau=2/\mu$ (only growth on one of the 2 sides) $c_{min}=a/\tau=r\mu$. The same minimum velocity appears at high negative values of drift, which have the effect of suppressing the contribution of diffusion. Positive values of drift instead transport the first bead freely ahead, and have the overall effect of a front moving, on average, at the converging speed $c=c_0+v$.

\begin{acknowledgments}
This work was financially supported by the Stichting voor Fundamenteel Onderzoek der Materie (FOM). We aknowledge Roberto Benzi, David Nelson and Tom de Greef for helpful discussions and Wencke Adriaens for help with handling of the bacteria. Partial support from the COST Action MP1305 is kindly acknowledged. We thank the technical support of Jorgen van der Veen and Henny Manders on the realization of the experimental set-up.
\end{acknowledgments}

\bibliography{multichannel_arxiv}

\clearpage
\onecolumngrid

\begin{center}
\textbf{\large Supporting Information (SI): Finite-size effects on bacterial population expansion under controlled flow conditions}
\end{center}

\setcounter{figure}{0}
\setcounter{page}{1}
\makeatletter
\renewcommand{\thefigure}{S\arabic{figure}}
\renewcommand{\bibnumfmt}[1]{[S#1]}
\renewcommand{\citenumfont}[1]{S#1}

Details on the experimental set-up, on the flow characterisation and on the procedure to extract the front speed of the bacteria colony from the intensity profiles are described here.

\begin{figure}[tbhp!]
\centering
\subfloat[]{\includegraphics[width=.35\linewidth]{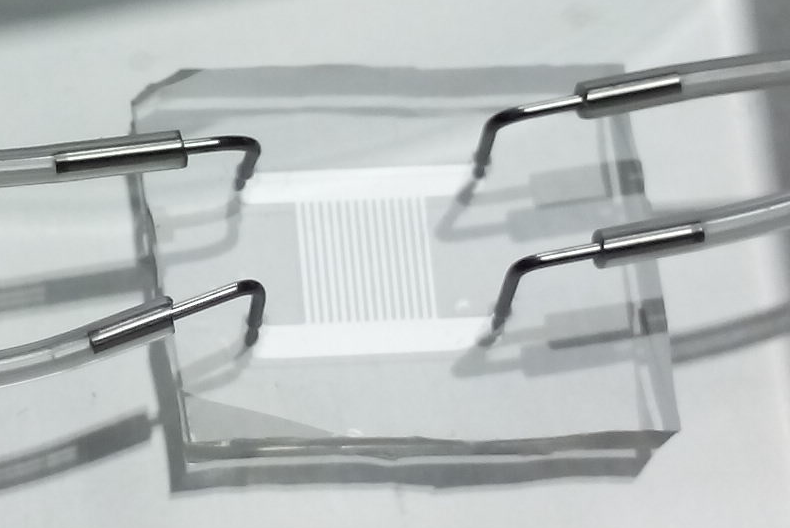}\label{device}}
\subfloat[]{\includegraphics[width=.3\linewidth]{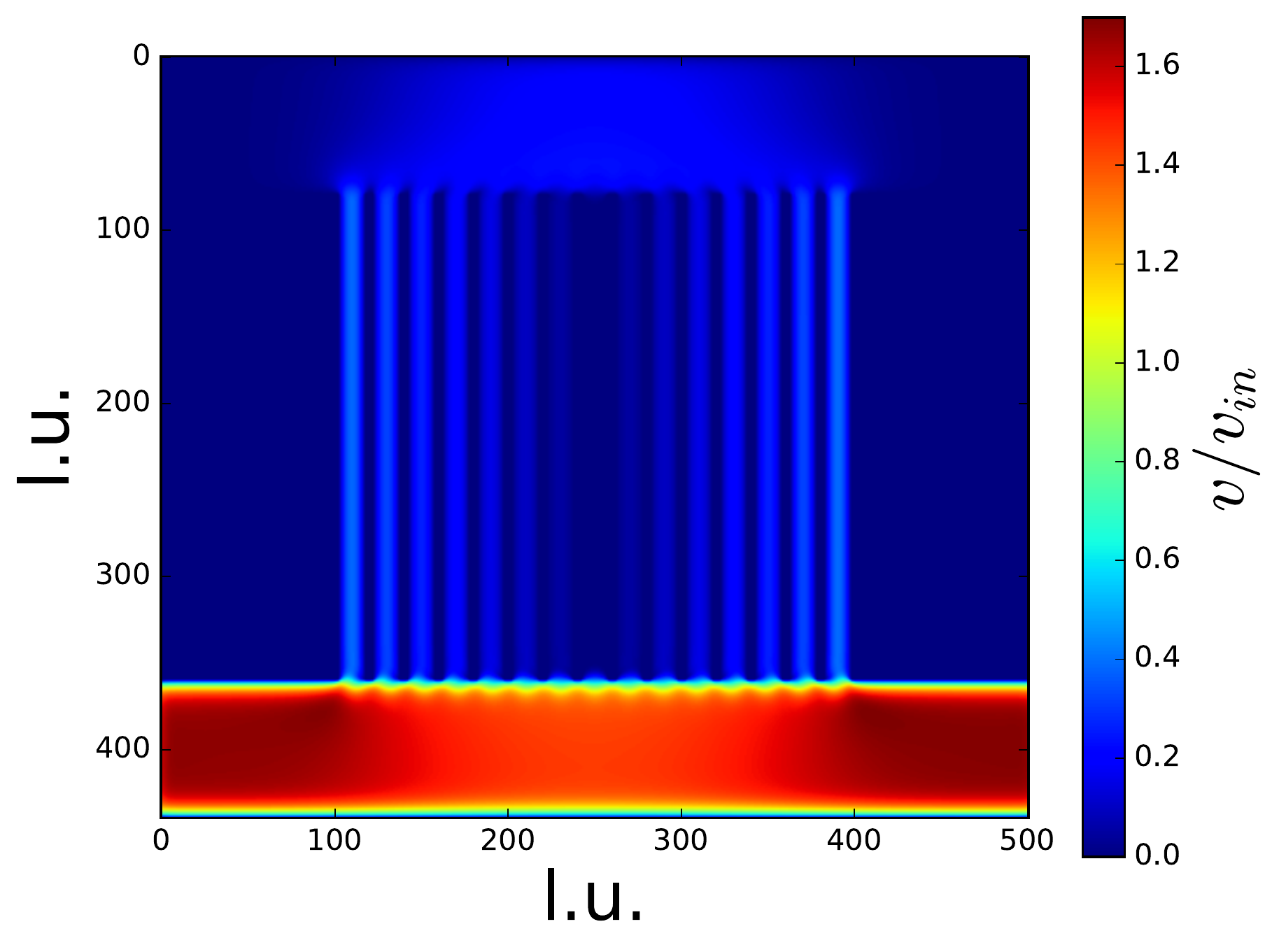}\label{LB}}
\caption{(a) A picture of a microfluidic device composed of $15$ channels with metal connectors for the external tubing system. (b) The dimensionless magnitude of the velocity in the mid-plane of the device obtained by LBM simulations: $v=\sqrt{v_x^2+v_y^2+v_z^2}$ normalised by the mean inflow velocity $v_{in}$. Size of the system is (501 $\times$ 440) l.u., with 1 l.u. = 20 $\mu$m.}

\end{figure}

\subsection*{Flow characterization}

\begin{figure}[tbhp!]
\centering
\subfloat[]{\includegraphics[width=.4\linewidth]{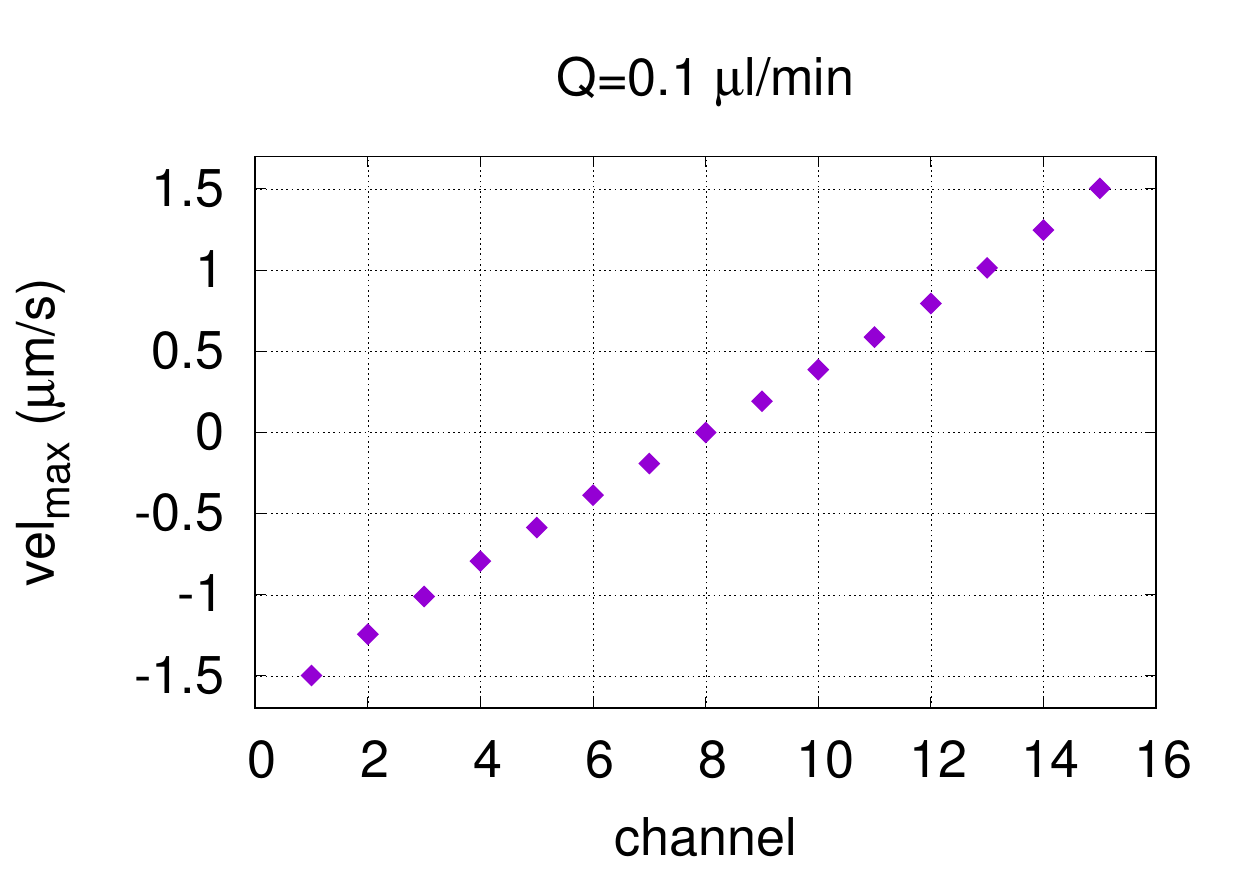}\label{LBQ}}
\subfloat[]{\includegraphics[width=.35\linewidth]{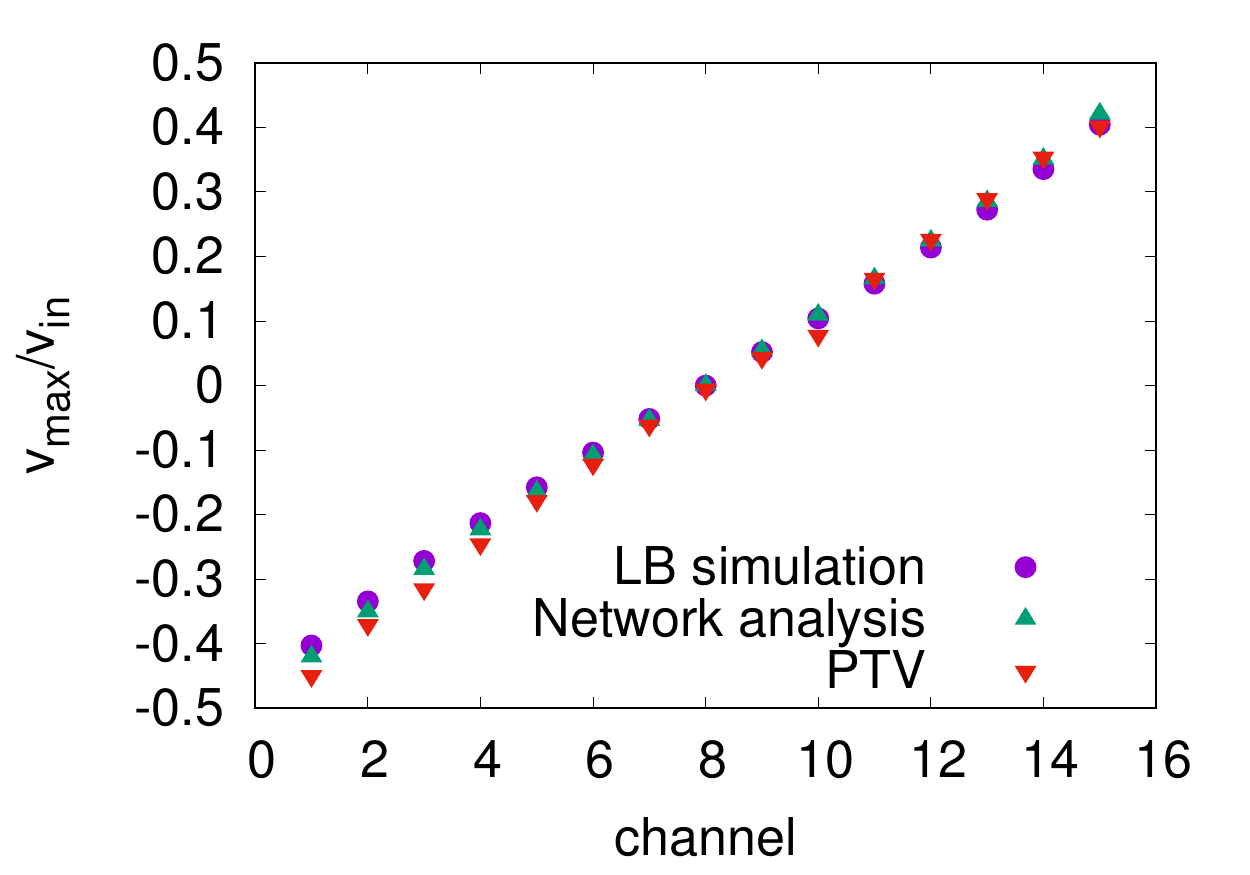}\label{comparison}}
\caption{(a) Maximum centreline velocity in the 15 channels with inflow rate Q=0.1$~\mu$L/min simulated with the Lattice Boltzmann Method. (b) Maximum velocity in the channels of the $15ch$ device normalised by the mean inflow velocity $v_{in}$. Comparison between the values obtained from a Lattice Boltzmann Simulation, the network analysis and PTV experiments on polystyrene beads.}
\end{figure}

The values of the imposed flow velocity in each channel have been obtained from a Lattice Boltzmann simulation. The $15ch$ device has been discretized into (501$\times$440$\times$14) l.u., which corresponds to 20 $\mu$m/l.u., while the $42ch$ device into (950$\times$475$\times$7) l.u., which means 40 $\mu$m/l.u.. The magnitude of the velocity field at mid-height in the device is shown in Figure \ref{LB}, where the data are normalised by the mean inflow velocity $v_{in}$. The simulations have been initialised by setting no-slip boundary conditions at the walls and inflow and outflow boundary conditions at the entrance and exit planes of the device. The estimated centreline velocity in the channels in the $15ch$ device is also plotted in Figure \ref{LBQ} for the typical inflow rate Q=0.1 $\mu$L/min.
This flow has been also validated both by using the network analysis, fixing the input flow rate and solving the pressure drop at each fluidic node, and in the device using particle tracking velocimetry (PTV) on polystyrene beads of radius 1.5 $\mu$m (PolyspherexTM). The PTV experiment has been performed at inflow rate Q=1 $\mu$L/min and the output has been analysed using \emph{Trackpy v0.3.0} \citeNew{allan_2015_34028}. The resulting velocities are normalised by the inflow velocity in order to compare them with the other methods (see Figure \ref{comparison}).

\subsection*{Image Processing}

The raw images acquired with fluorescent filters are processed and analysed with Python scripts. As general procedure, the noise outliers pixels are removed in the whole time stack, by replacing them with the median 2-pixel radius filter value, then the first image, characterised by no bacteria in the channels is subtracted to all images as background and rotation is applied, if needed.  
\begin{figure}[h!]
\centering
\subfloat[]{\includegraphics[width=.4\linewidth]{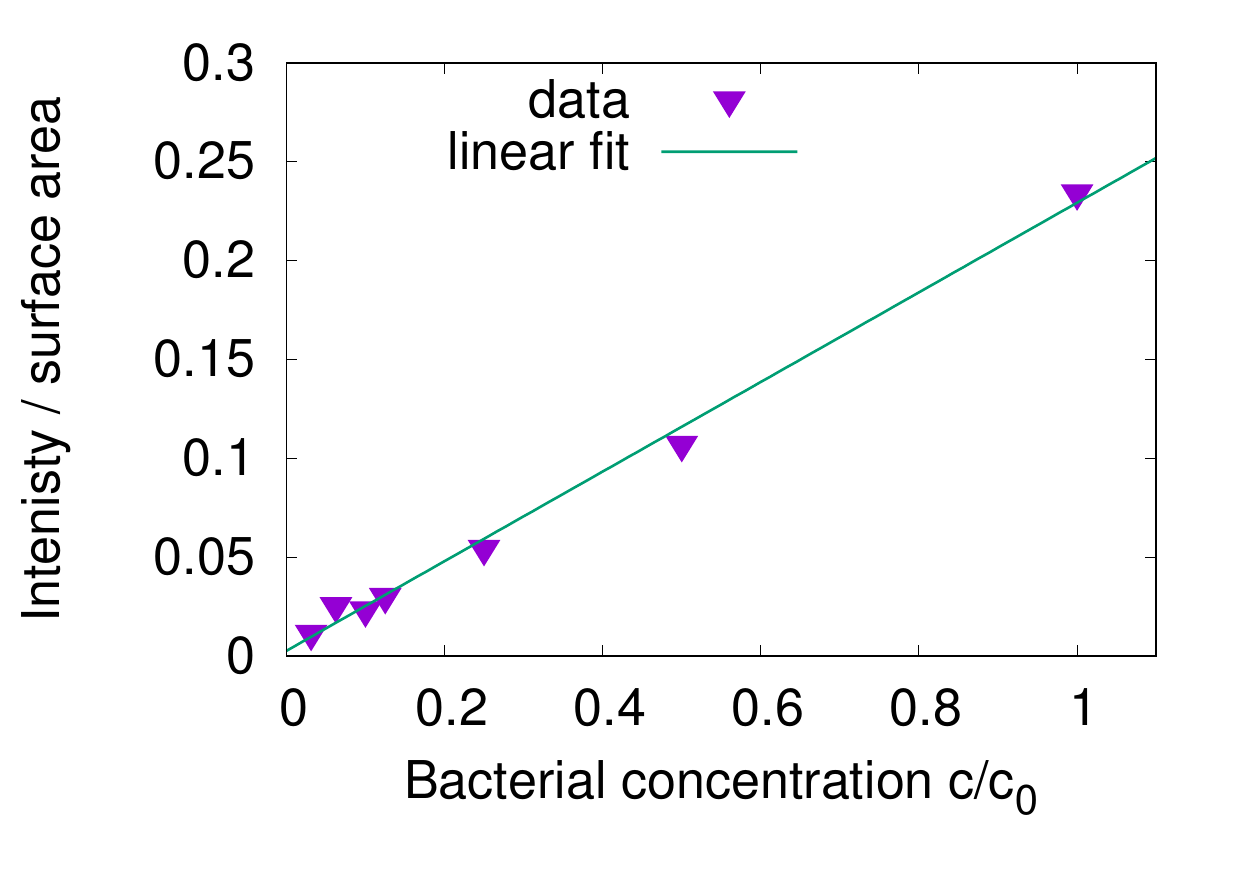}\label{proportionality}}\\
\subfloat[]{\includegraphics[width=.8\linewidth]{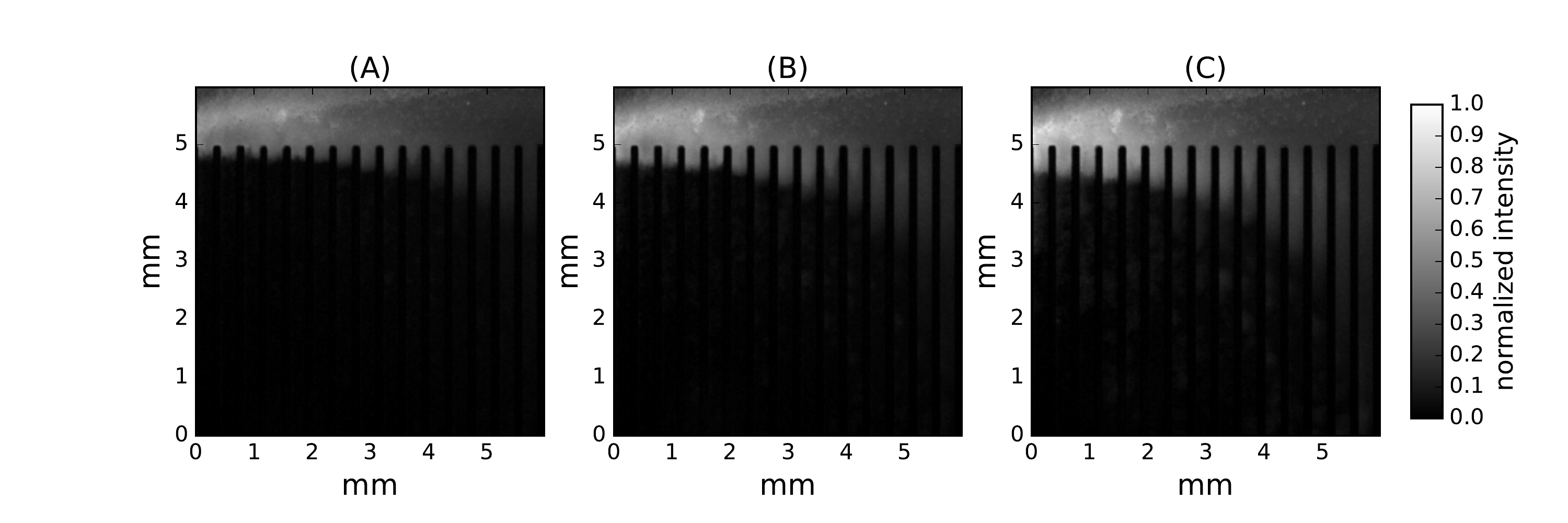}\label{3snapshots}}
\caption{(a) Proportionality between bacteria concentration and fluorescent intensity collected by the camera. (b) three snapshots of fluorescent intensity in the $15ch$ device for an experiment with inflow Q=0.1 $\mu$L/min, corresponding to time $t_1$=5.7, $t_2$=6.7 and $t=7.6$ hours from the beginning of the experiment. The data plotted here has also been filtered by a 3-pixel radius gaussian filter. The initial colony of bacteria deposited in the upper part of the device, grows along the channels experiencing different flow velocities.}
\end{figure}

\begin{figure}
\centering
\includegraphics[width=.7\linewidth]{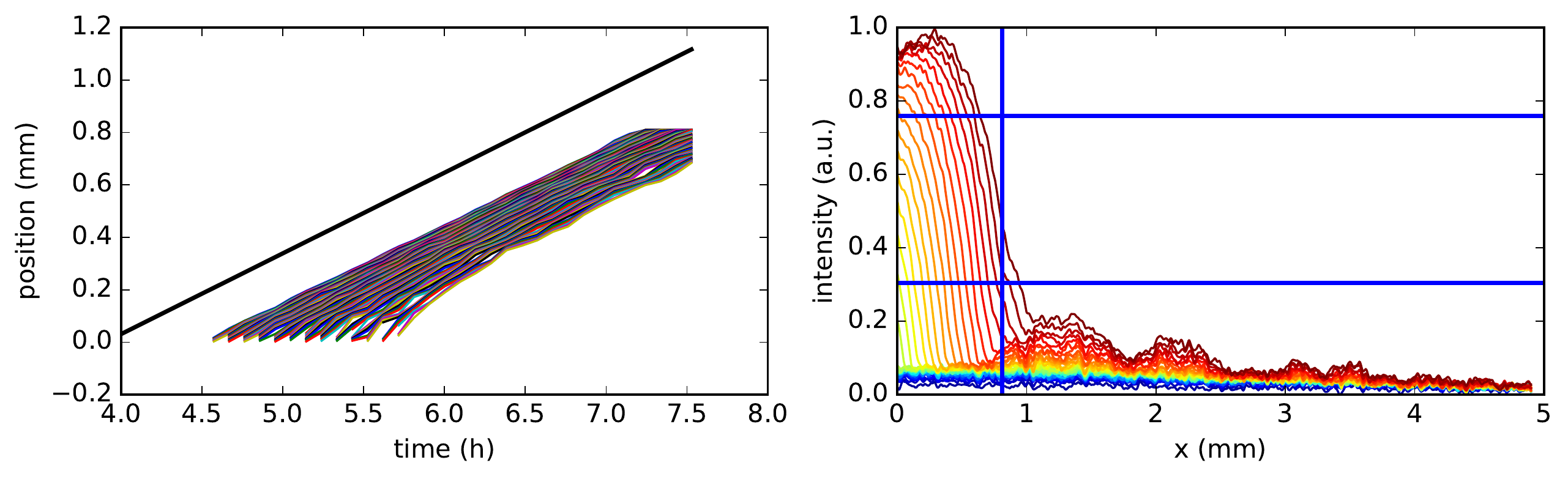}
\caption{To determine the front speed multiple thresholds, around 200, are considered between the minimum and the maximum level of signal intensity (2 horizontal lines on the right). For each choice of the threshold it corresponds, at every time value, a specific position of the front (values on the left). A regression line can be extracted for each series of data with its regression coefficient. The final weighted mean slope is plotted on the left as black line. The vertical line on the right plot corresponds to the cutoff on the signal. Only the signal at the left of the cutoff is considered.}
\label{method}
\end{figure}
Figure \ref{proportionality} shows the proportionality between the bacteria concentration, inside channels of cross section 300 $\mu$m $\times$ 280 $\mu$m, and the fluorescent intensity images collected by the camera and post-processed in the usual way. The points in the graph refer to the mean intensity per unit surface, measured from images of independent channels homogeneously filled by several dilutions of a reference bacteria concentration $c_0$. The x-axis spans a reasonably wide range of concentrations, similar to the ones of the experiment.
In Figure \ref{3snapshots}, three snapshots of the bacteria fluorescent intensity in the device are shown. The pixel intensity is usually integrated along the transversal direction of each channel in order to express the intensity profile only along the direction of growth, finally a Gaussian 3-pixel filter is applied to this one-dimensional signal to smooth it. At this stage we have a time series of bacteria density for each channel along the overall propagation direction. The front profile is usually disturbed by a background growth which at late time is covering the signal. This growth is not affecting the measurement if we consider the front only at its initial stage. Operatively, we measure the front only before a maximum time value in combination with a cutoff on the signal on the x-axis, as shown in Figure \ref{method} on the right. The front position can be determined by fixing a threshold on the signal intensity and its speed can be calculated as the slope of a regression line of the detected front positions over their times. In order to make the speed estimate independent on the threshold choice when the front shape is not regular, we proceed by fixing a set of incremental thresholds, between a minimum and maximum level, and averaging the corresponding set of slopes. This procedure is sketched in Figure \ref{method}. The mean slope, weighted by the linear correlation coefficient, which quantifies the quality of the linearity, is the extracted speed of the front and the standard deviation of the distribution of slopes is taken as the error. In this way bad choices of threshold are automatically neglected and more importance is given to the parts of the front which are propagating at constant speed. To measure the width of the fronts a fit with a hyperbolic tangent is performed on the raw front profiles as shown in Figure \ref{fit_tanh}.

\subsection*{Front width and maximum density}

\begin{figure}
\centering
\subfloat[]{\includegraphics[width=.3\linewidth]{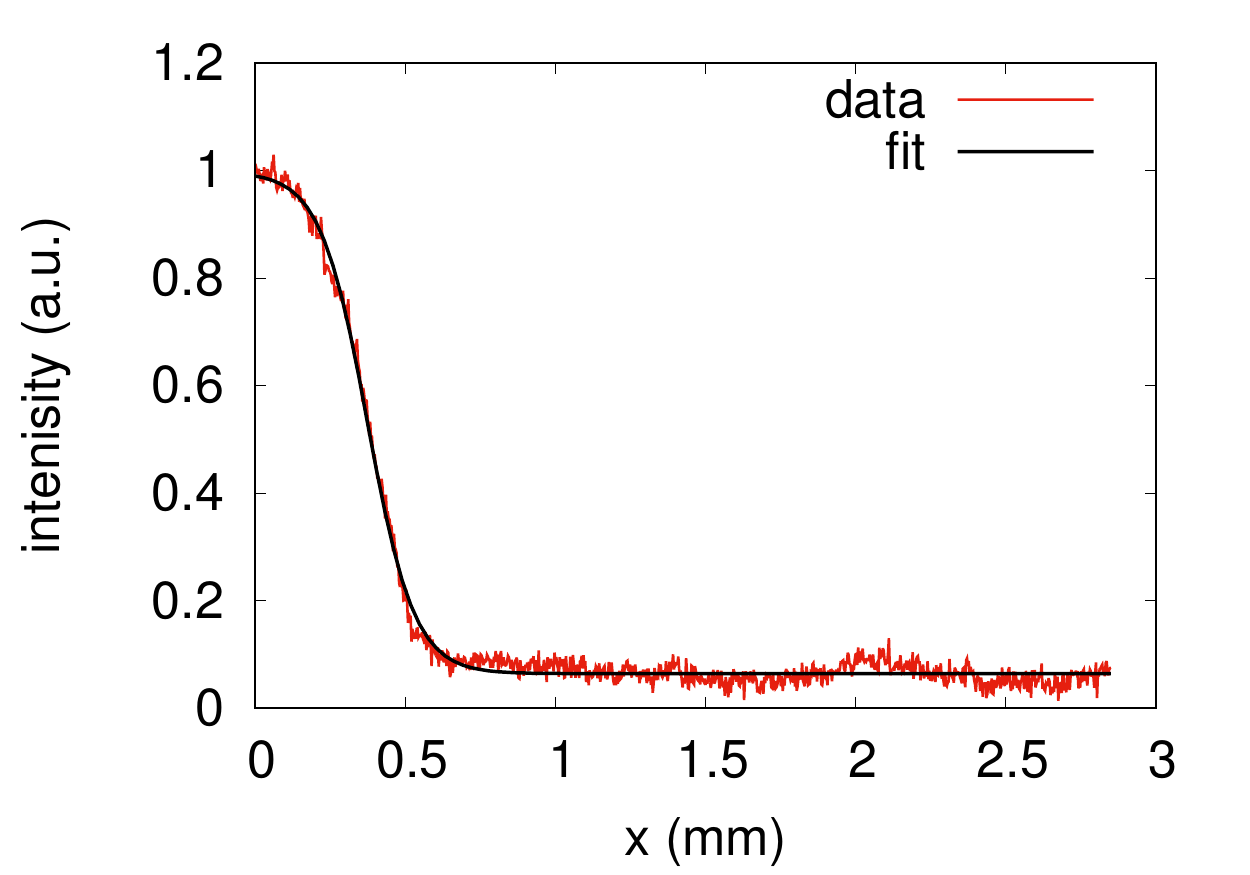}\label{fit_tanh}}
\subfloat[]{\includegraphics[width=.3\linewidth]{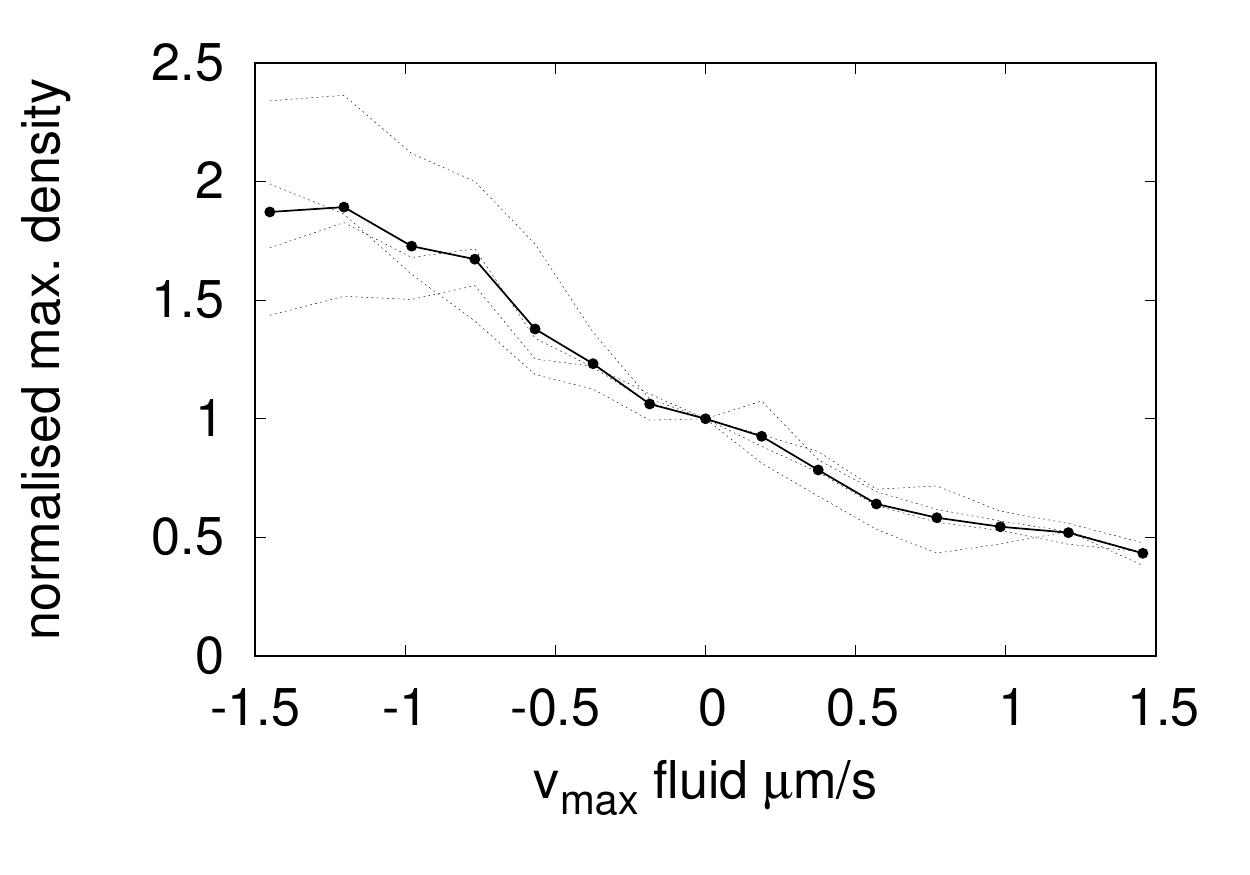}\label{carry}}
\subfloat[]{\includegraphics[width=.3\linewidth]{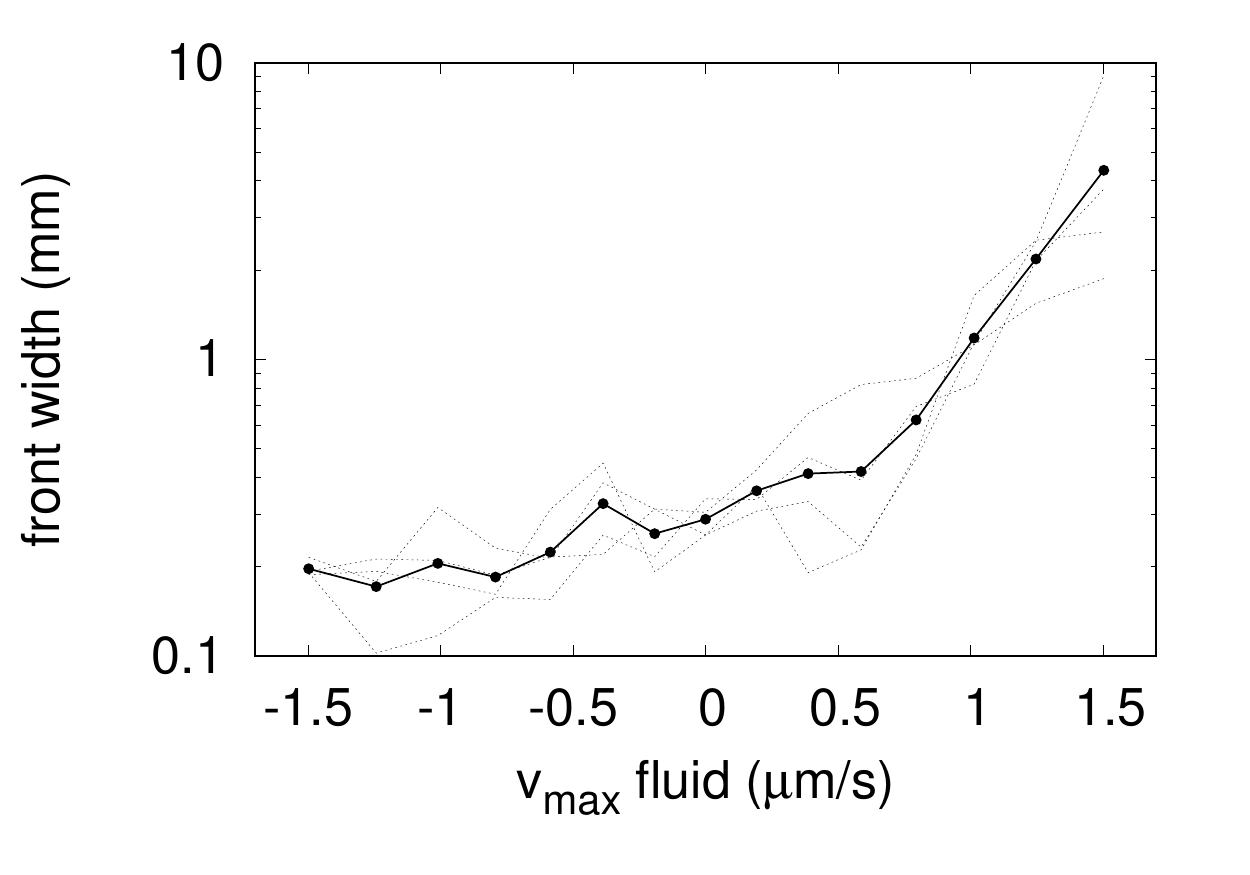}\label{width}}
\caption{(a) Fit of the front raw profile by a hyperbolic tangent on channel number 4. (b) Maximum density of bacteria behind the front in the stationary case as a function of maximum velocity in the channel, normalised to the maximum density in the zero flow channel. Black dots are the mean values of four repeated single experiments (gray lines), which give an idea of the uncertainty. (c) Front widths as a function of maximum velocity in the channel plotted in a logarithmic scale on y-axis. Gray lines refer to 4 identical experiments, while the black dots are the mean values.}
\label{extra}
\end{figure}
We quantify the intensity profiles in the channels corresponding to the experiments with Q=0.1 $\mu$L/min and integrate the fluorescent intensity in the part of the channel behind the front in the stationary case. These values are plotted for the four identical experiments as a function of the maximum velocity in the channels in Fig. \ref{carry} together with the mean value. We observe that the density is systematically higher for negative velocities, while the populations appears systematically diluted in the co-flow regime. Finally, also the front width has been estimated from the front profiles. The data come from a fit on the front profiles with an hyperbolic tangent function, $f(x)=a[1-b\tanh{(cx+d)}]$, and the width corresponds to $2/c$. The fits have been performed at different times and the front width has been extracted at the stationary regime, which can also corresponds to slightly different time values. The width shown in Figure \ref{width} appears to have a systematic variation with the flow direction and intensity, in particular to be larger in the co-flow regime.

\subsection*{1D hard beads model}
Numerical data from the one-dimensional model have been obtained from the average of $20$ independent realisations. For every realisation the front speed has been calculated over 1200 generation time or, for strong counter-drift, over a length of around $1200r$. Parameters of this simulations are $D=0.0036$, $\mu=0.5$, $r=0.088$ and $dt=0.05$. The simulation is initialised always with three beads.

\end{document}